\documentclass[twocolumn,english,superscriptaddress,citeautoscript,showpacs,preprintnumbers,amsmath,amssymb,prd,floatfix,nofootinbib]{revtex4-1}
\pdfoutput=1
\usepackage{amsmath, amsthm, amssymb, graphicx,bm,amstext,subfigure, mathdots}
\usepackage{hyperref}
\usepackage{color}

\begin{document}
\title{How the cosmological constant is hidden by Planck scale curvature fluctuations}
\author{Qingdi Wang}
\author{William G. Unruh}
\affiliation{Department of Physics and Astronomy, 
The University of British Columbia,
Vancouver, Canada V6T 1Z1}
\begin{abstract}
It is argued in a recent letter \cite{PhysRevLett.123.131302} that the effect of a large cosmological constant can be naturally hidden in Planck scale curvature fluctuations. We point out that there are problems with the author's arguments. The hiding of the cosmological constant proposed in \cite{PhysRevLett.123.131302} by choosing a suitable lapse function is just an illusion maintained by external forces. In particular, it can not be achieved if the cosmological constant is positive. Fortunately, it works for a negative cosmological constant in a different way, and, interestingly, the sign of the cosmological constant just needs to be negative to make the average spatial curvature $\langle R\rangle$ small.
\end{abstract}
\maketitle

\section{Introduction}
The cosmological constant problem is a long standing problem in mordern physics. The huge vacuum energy is usually expected to produce a large cosmological constant which leads to a disastrous gravitaional effect. In a recent letter \cite{PhysRevLett.123.131302} the author argues that the fluctuations in the metric at Planck scales (Wheeler's spactime foam) make it possible to hide the effect of a large cosmological constant.

We certainly agree with the author of \cite{PhysRevLett.123.131302} that the fluctuations in the metric must be taken into account, and have previously suggested how this might come about \cite{PhysRevD.95.103504, Qingdi:2019, Wang:2019mee}. Unfortunately the author's proposal suffers from some problems. In this paper, we first show in Sec.\ref{comment} that the above arguments have problems and thus the hiding actually does not work in the way proposed in \cite{PhysRevLett.123.131302}. We then investigate whether it is possible to make the idea of hiding the cosmological constant in Planck scale curvature fluctuations work in other ways. We show in Sec.\ref{positive lambda} that this idea does not work for a positive cosmological constant due to the universal divergences of the geodesics. The small scale spacetime fluctuations do not help in this situation. Fortunately, this idea works for a negative cosmological constant in a different way. It is interesting that the sign of the cosmological constant just needs to be negative to make the average spatial curvature $\langle R\rangle$ small. We show this different way of hiding the cosmological constant in Sec.\ref{negative lambda1}. 

\section{Problems with the author's arguments}\label{comment}
The author of \cite{PhysRevLett.123.131302} employs the initial value formulation of general relativity and takes the shift vector to be zero for simplicity. This is essentially assuming the metric of the form
\begin{equation}\label{metric}
ds^2=-N^2dt^2+g_{ij}dx^idx^j.
\end{equation}
He considers the volume averaging on the initial hypersurface $t=0$
\begin{equation}\label{definition}
\left\langle X\right\rangle_{\mathcal{U}}=\frac{1}{V_\mathcal{U}}\int_{\mathcal{U}}X\sqrt{g}d^3x\quad\text{with}\quad V_{\mathcal{U}}=\int_{\mathcal{U}}\sqrt{g}d^3x,
\end{equation}
where the region $\mathcal{U}$ is defined in some time-independent way. 

It is argued that a large class of initial data on the hypersurface $t=0$ can exhibit zero average expansion $\langle K\rangle=0$. It is further argued that the classical time evolution can preserve this property since one can choose a suitable lapse function $N$ to make $d^n\langle K\rangle/dt^n=0$ for all $n>0$.

More concretely, the author uses the equation for the rate of change of $K$
\begin{eqnarray}\label{K derivative equation}
\frac{dK}{dt}&=&N\left(-K^2-R+3\Lambda\right)+D^iD_iN\nonumber\\
&=&N\left(-\frac{K^2}{3}-2\sigma^2+\Lambda\right)+D^iD_iN
\end{eqnarray}
and the relation
\begin{equation}
\frac{d\sqrt{g}}{dt}=NK\sqrt{g}
\end{equation}
to obtain the rate of change of the average expansion $\langle K\rangle$ with respect to the coordinate time $t$ :
\begin{eqnarray}\label{correct}
\frac{d\langle K\rangle}{dt}=&&\frac{1}{V_{\mathcal{U}}}\int_{\mathcal{U}}N\left(-R+3\Lambda+\frac{D^iD_iN}{N}\right)\sqrt{g}d^3x\\
=&&\frac{1}{V_{\mathcal{U}}}\int_{\mathcal{U}}N\left(\frac{2K^2}{3}-2\sigma^2+\Lambda+\frac{D^iD_iN}{N}\right)\sqrt{g}d^3x.\nonumber
\end{eqnarray}
The corresponding Eq.(7) in \cite{PhysRevLett.123.131302} omitted the term $D^iD_iN$ since it is a total derivative that reduces to a surface integral. We keep this term in \eqref{correct} since it is not necessarily to be zero after integration.

It is then argued that since the integrand in \eqref{correct} doesn't have a definite sign, there will be infinite choices of $N$ for which the right-hand side of \eqref{correct} vanishes. Similar arguments are also made for higher order time derivatives of $\langle K\rangle$. In this way, the author finds a foliation of spacetime by slices of vanishing average expansion and then concludes that the effect of the large cosmological constant is nearly invisible at observable scales.

Unfortunately there are problems with the above arguments. In fact, a choice of lapse corresponds to a choice of coordinates, and no physics can depend purely on the choice of coordinates. If distances between geodesics, or more importantly, the wavelengths of fields, grow with time (usually proper time, not coordinate time) they will do so in all coordinate systems, and cannot be hidden by a coordinate choice.

As a counterexample we can look at the de Sitter space which is a vacuum solution of the Einstein equations with a positive cosmological constant $\Lambda$. In this spacetime one can choose the static slicing coordinate
\begin{equation}\label{static coordinate}
ds^2=-\left(1-\frac{\Lambda}{3}r^2\right)dt^2+\frac{1}{1-\frac{\Lambda}{3}r^2}dr^2+r^2d\Omega^2.
\end{equation}
The spatial slices $t=Constants$ of \eqref{static coordinate} have expansion $K\equiv 0$ but physically the de Sitter spacetime is exponentially expanding. This exponential expansion can be seen by transforming the static coordinate \eqref{static coordinate} to the following flat slicing coordinate (FLRW)
\begin{equation}\label{flat slicing}
ds^2=-d\tau^2+e^{2\sqrt{\frac{\Lambda}{3}}\tau}\left(dx^2+dy^2+dz^2\right).
\end{equation}

The lesson learned from this counterexample is that we should not choose the lapse function $N$ arbitrarily. In fact, $K$ is the local volume expansion rate perceived by the stationary observers defined by $x^i=Constants$ (Eulerian observers). In the static slicing \eqref{static coordinate} $N$ is position dependent, $x^i=Constants$ are not geodesics, so that these observers are accelerating. There are external forces acting on them to maintain their constant spatial positions. In the flat slicing \eqref{flat slicing} $N=1$, $x^i=Constants$ are geodesics so that these observers are free falling. The expansion $K\equiv 0$ in \eqref{static coordinate} because the gravitational repulsions caused by the positive $\Lambda$ are balanced by the external forces, it does not mean the effect of $\Lambda$ is invisible.

Therefore, we should use free falling observers who only feel gravity to test physically whether the space is expanding or contracting. Technically, the acceleration of the stationary observer is tangent to the hypersurfaces $t=Constants$ with the $i$th component of the accelearation given by $a_i=D_iN/N$ (see Eq.(3.17) in \cite{Gourgoulhon:2007ue}). So the lapse function $N$ should be chosen to be spatially independent to make sure $x^i=Constants$ are geodesics. In this case, the rate of change of the average expansion $\langle K\rangle$ perceived by these free falling observers given by \eqref{correct} is
\begin{equation}\label{geodesic average}
\frac{d\langle K\rangle}{d\tau}=3\Lambda,
\end{equation}
where $\tau=\int Ndt$ is the proper time of these observers and we have used the requirement that the average spatial curvature $\langle R\rangle=0$.

On the initial hypersurface $\Sigma$, these free falling observers have the same unit tangent vectors with the Eulerian observers defined by $x^i=Constants$ when the lapse $N$ is position dependent, i.e., they have the same initial velocities. The only difference is that they have different accelerations---the free falling observers have zero accelerations while the Eulerian observers have accelerations $a_i=D_iN/N$ produced by the external forces. We see from \eqref{geodesic average} that the free falling observers still see the effect of the cosmological constant, the hiding of the cosmological constant seen by the Eulerian observers is just an illusion maintained by the external forces. This result is quite natural since one should not try to hide the cosmological constant by choosing $N$ in the first place.

Moreover, even for the non-inertial Eulerian observers, $d\langle K\rangle/dt$ is not a physical quantity observed by them. One should use the Eulerian observers' proper times $\tau$ instead of the coordinate time $t$. In fact, the infinitesimal local volume element observed by each Eulerian observer is $\sqrt{g}d^3x$. The rate of change of $\sqrt{g}$ and the rate of change of $d\sqrt{g}/d\tau$ perceived by each Eulerian observer are
\begin{eqnarray}
\frac{d\sqrt{g}}{d\tau}&=&K\sqrt{g},\\
\frac{d^2\sqrt{g}}{d\tau^2}&=&\left(-R+3\Lambda+\frac{D^iD_iN}{N}\right)\sqrt{g}.
\end{eqnarray}
Note that the proper times $\tau$ are different from point to point.

The quantities $\sqrt{g}$, $d\sqrt{g}/d\tau$ and $d^2\sqrt{g}/d\tau^2$ are physical quantities that actually observed by each Eulerian observer. Integrating $\sqrt{g}$ over the region $\mathcal{U}$ gives
\begin{equation}
V_{\mathcal{U}}=\int_{\mathcal{U}}\sqrt{g}d^3x,
\end{equation}
which is just the macroscopic volume defined by the second equation in \eqref{definition}. Integrating $d\sqrt{g}/d\tau$ over $\mathcal{U}$ and then divide the volume $V_{\mathcal{U}}$ gives the average of $d\sqrt{g}/d\tau$:
\begin{equation}
\overline{\frac{d\sqrt{g}}{d\tau}}=\frac{1}{V_{\mathcal{U}}}\int_{\mathcal{U}}K\sqrt{g}d^3x,
\end{equation}
which is just the average expansion $\langle K\rangle$ defined by the first equation in \eqref{definition}. 

However, integrating $d^2\sqrt{g}/d\tau^2$ over $\mathcal{U}$ and then divide the volume $V_{\mathcal{U}}$ gives the average of $d^2\sqrt{g}/d\tau^2$:
\begin{eqnarray}\label{physical effect}
\overline{\frac{d^2\sqrt{g}}{d\tau^2}}&=&\frac{1}{V_{\mathcal{U}}}\int_{\mathcal{U}}\left(-R+3\Lambda+\frac{D^iD_iN}{N}\right)\sqrt{g}d^3x\nonumber\\
&=&3\Lambda+\frac{1}{V_{\mathcal{U}}}\int_{\mathcal{U}}\frac{D^iD_iN}{N}\sqrt{g}d^3x,
\end{eqnarray}
where we have used the requirement that the average spatial curvature $\langle R\rangle=0$ in obtaining the second line of \eqref{physical effect}. The integration of the term $D^iD_iN/N$ in \eqref{physical effect} represents the average effect of the external forces acting on the Eulerian observers.

Comparing to \eqref{correct}, the above expression \eqref{physical effect} does not have the factor $N$ in the integrand and since $N$ is position dependent as supposed in \cite{PhysRevLett.123.131302}, we would have
\begin{equation}
\frac{d\langle K\rangle}{dt}\neq\overline{N}\overline{\frac{d^2\sqrt{g}}{d\tau^2}}.
\end{equation}
In other words, the unphysical quantity $d\langle K\rangle/dt$ is in general different from the physical quantity $\overline{d^2\sqrt{g}/d\tau^2}$ and one can not choose $N$ in the way proposed in \cite{PhysRevLett.123.131302} to make $\overline{d^2\sqrt{g}/d\tau^2}=0$. One may still choose $N$ in different ways to make $\overline{d^2\sqrt{g}/d\tau^2}=0$, for example, one can choose $N$ to be the eigenfunction of the Laplace operator corresponding to the eigenvalue $3\Lambda$, i.e., $-D^iD_iN=3\Lambda N$. However, again, this choice of $N$ is just a coordinate choice, the hiding of $\Lambda$ in this way is just an illusion maintained by the external force.

In summary, we have shown that the author's argument is problamatic. As a result, the hiding does not work in the way proposed in \cite{PhysRevLett.123.131302}. Does it work in other ways? Further investigations will be given in the following sections.

\section{$\Lambda>0$ does not work}\label{positive lambda}
In this section we show from a different perspective that the inhomogeneous Planck scale curvature fluctuations can not hide the effect of a positive $\Lambda$.

Consider a free falling observer $\gamma$ in a spacetime with $\Lambda>0$. The dynamics of an infinitesimally nearby free falling test particle observed in $\gamma$'s own local inertial frame is given by the geodesic deviation equation (see e.g. pages 47, 225 of \cite{Wald:1984rg}):
\begin{equation}\label{geodesic deviation1}
\frac{d^2\xi^i}{d\tau^2}=-\sum_{j=1}^3  R^i_{0j0}(\tau)\xi^j, \quad i=1, 2, 3,
\end{equation}
where $\tau$ is $\gamma$'s proper time, $\xi^i$ is the coordinate of the deviation vector from $\gamma$ to the test particle in $\gamma$'s local inertial frame, $R^i_{0j0}$ are components of the Riemann curvature tensor along $\gamma$.

The Riemann tensor can be expressed in terms of the Weyl tensor and the Ricci tensor:
\begin{equation}
R^a_{bcd}=C^a_{bcd}+\delta^a_{[c}R_{d]b}-g_{b[c}R^a_{d]}-\frac{1}{3}R\delta^a_{[c}g_{d]b}.
\end{equation}
In $\gamma$'s own local inertial frame the metric components $g_{\mu\nu}$ along $\gamma$ is exactly $\eta_{\mu\nu}=\text{diag}(-1, 1, 1, 1)$ so that we have
\begin{equation}\label{Weyl}
R^i_{0j0}=C^i_{0j0}-\frac{1}{2}R^i_j+\frac{1}{2}\delta^i_jR_{00}+\frac{1}{6}R\delta^i_j.
\end{equation}

The Ricci tensor is determined by the Einstein equations:
\begin{equation}\label{EFE}
R_{ab}=\Lambda g_{ab}.
\end{equation}
Plugging \eqref{EFE} into \eqref{Weyl} gives
\begin{equation}
R^i_{0j0}=C^i_{0j0}-\frac{\Lambda}{3}\delta^i_j.
\end{equation}
Then the geodesic deviation equation \eqref{geodesic deviation1} can be written as
\begin{equation}\label{geodesic deviation}
\frac{d^2\mathbf{x}}{d\tau^2}=\left(\frac{\Lambda}{3}I-C\right)\mathbf{x},
\end{equation}
where $\mathbf{x}=(\xi^1, \xi^2, \xi^3)^t$, $I=\mathrm{diag}(1, 1, 1)$ is the identity matrix, $C=(C^i_{0j0})_{3\times 3}$ is a matrix whose elements $C^i_{0j0}$ are components of the Weyl tensor along the world line of $\gamma$. One important property of the Weyl tensor is that it is trace free:
\begin{equation}\label{trace free}
C^a_{0a0}=\sum_{i=1}^3C^i_{0i0}=0,
\end{equation}
where we have used $C^0_{000}=0$ which is required by the symmetry property of the Weyl tensor.
 
The Planck scale curvature fluctuations are encoded in the Weyl tensor. These fluctuations are inhomogeneous and anisotropic. However, the statistical properties of these fluctuations should still be homogeneous and isotropic, i.e., the observer $\gamma$ should see the same magnitude of fluctuations in every point and in every direction. Thus we would have the expectation values of the off-diagonal components
\begin{equation}
\left\langle C^i_{0j0}\right\rangle=0,\quad i\neq j,
\end{equation}
and the diagonal components
\begin{equation}
\langle C^1_{010}\rangle=\langle C^2_{020}\rangle=\langle C^3_{030}\rangle. \label{equal diagonal component}
\end{equation}
Taking expectation values on both sides of \eqref{trace free} and use the property \eqref{equal diagonal component} we obtain that the diagonal components $\left\langle C^i_{0i0}\right\rangle=0$. Thus the Weyl tensor term $C$ in \eqref{geodesic deviation} fluctuates around $0$ and provides a fluctuating tidal force on the test particle. On average the test particle would move along a smooth path driven by the cosmological constant term $\Lambda/3$ and at the same time execute oscillations around this path due to the fluctuations of the Weyl tensor term $C$. This averaged smooth path is given by the solution of \eqref{geodesic deviation} when the fluctuation term $C$ is excluded:
\begin{equation}\label{solution}
\bar{\xi}^i= c_ie^{\sqrt{\frac{\Lambda}{3}}\tau}+c'_ie^{-\sqrt{\frac{\Lambda}{3}}\tau}, \quad i=1, 2, 3,
\end{equation}
where $c_i$ and $c'_i$ are integration constants. The constant $c_i$ is zero only for the very special case when initially the test particle is moving toward $\gamma$ with a speed $\frac{d\bar{\xi}^i}{d\tau}(0)=-\sqrt{\frac{\Lambda}{3}}\bar{\xi}^i(0)$. Consider the perpetual perturbation from the fluctuations of the Weyl tensor term $C$, this special initial condition is impossible to be satisfied so that $c_i$ must be nonzero. Then the first term in \eqref{solution} would quickly become dominant that the averaged smooth path goes as
\begin{equation}
\bar{\xi}^i\sim c_ie^{\sqrt{\frac{\Lambda}{3}}\tau}, \quad i=1, 2, 3.
\end{equation}
So repulsive force produced by the positive $\Lambda$ would accelerate the nearby test particle away from $\gamma$ exponentially fast. The Planck scale curvature fluctuations encoded in $C$ make the test particle oscillate around this exponential path. 

The deviation vector describes how the infinitesimal distances between neighboring geodesics evolve with time. The distances between two far away geodesics in a geodesic congruence can be obtained by integrating these infinitesimal distances. Of course there are ambiguities in doing the integration since, in curved spacetime, the spatial distance is only well defined for infinitesimal distances. There is no unique definition for large spatial distances. However, since on average all the infinitesimal distances grow exponentially, any sensible definition of the integration would give, on average, exponential growth between large-distance geodesics.

In other words, consider a macroscopic ball of free falling test particles, each particle in this ball would be wildly fluctuating in response to the Weyl curvature fluctuations, and the average distance between any two nearby particles would finally be exponentially increasing. Since this average distance increasing is universal for any two neighboring particles, the volume of the macroscopic ball must also be exponentially increasing. This means that the effect of a positive $\Lambda$ can not be hidden in Planck scale curvature fluctuations---the spacetime would still explode.

\section{$\Lambda<0$ works}\label{negative lambda1}
It seems that from \eqref{physical effect} the hiding does not work no matter the sign of $\Lambda$. We have also shown in the last section that it is impossible for $\Lambda>0$ to work by a more general proof. Fortunately, this is not the end of the story. $\Lambda<0$ may work in a different way.

Define the local scale factor $a$ which describes the local ``size" of space by $g=a^6$, then the expansion $K=\frac{3}{a}\frac{da}{d\tau}$ and Eq.\eqref{K derivative equation} becomes
\begin{equation}\label{evolution equation}
\frac{d^2a}{d\tau^2}+\frac{1}{3}\left(2\sigma^2-\Lambda-\frac{D^iD_iN}{N}\right)a=0.
\end{equation}
As discussed in Sec.II that the lapse function $N$ needs to be spatially independent or at least the average of the term $D^iD_iN/N$ needs to be zero. Then since we always have $2\sigma^2-\Lambda>0$, $a$ must oscillate around $0$. Every time when $a$ crosses $0$, $K$ jumps discontinuously from $-\infty$ to $+\infty$. Similar to the derivative of the step function who jumps from $0$ to $1$ is a $\delta$ function, $dK/d\tau$ at $a=0$ is also a $\delta$ function:
\begin{equation}
\frac{dK}{d\tau}|_{a=0}=\mu\delta(a),
\end{equation}
where $\mu=+\infty$ because $K$ jumps from $-\infty$ to $+\infty$. Then we have
\begin{equation}
\frac{d^2\sqrt{g}}{d\tau^2}|_{a=0}=\mu\delta(a)\sqrt{g}
\end{equation}
and \eqref{physical effect} becomes
\begin{eqnarray}\label{correct average}
\overline{\frac{d^2\sqrt{g}}{d\tau^2}}=\frac{1}{V_{\mathcal{U}}}\Bigg(&&\int_{\mathcal{U}\cap \{a\neq 0\}}\left(-R+3\Lambda+\frac{D^iD_iN}{N}\right)\sqrt{g}d^3x\nonumber\\
+&&\int_{\mathcal{U}\cap \{a=0\}}\mu\delta(a)\sqrt{g}d^3x\Bigg).
\end{eqnarray}
The first term on the right hand side of \eqref{correct average} is negative while the second term is positive. They can cancel each other to make $\overline{\frac{d^2\sqrt{g}}{d\tau^2}}=0$ if the average oscillation amplitude of $a$ is a constant.

In this picture, $a=0$ are actually curvature singularities. It has been proved that the singularities must occur for a globally hyperbolic vacuum spacetime with a negative cosmological constant \cite{1976ApJ...209...12T}. Physically, the negative cosmological constant produces attractive effect which makes $\frac{d^2\sqrt{g}}{d\tau^2}<0$ (the first term on the right hand side of \eqref{correct average}) at points away from the singularities, and at the singularities bounces happen which produces repulsive effect which makes $\frac{d^2\sqrt{g}}{d\tau^2}>0$ (the second term on the right hand side of \eqref{correct average}). Macroscopically, the attractions at points away from the singularities are balanced by the the repulsions at the singularities. In this way, the effect of a large negative cosmological constant can be hidden in Planck scale curvature fluctuations. On the contrary, the effect of a large positive cosmological constant can not be hidden because a positive $\Lambda$ always produce repulsive effects, no mechanisms to produce attractive effects to balance the repulsiveness.

In addition, the sign of the cosmological constant just needs to be negative to make the average spatial curvature $\langle R\rangle$ small. In fact, taking average on Eq.(1a) of \cite{PhysRevLett.123.131302} we have
\begin{equation}\label{average constraint}
\left\langle R\right\rangle=2\Lambda+\left\langle K_{ij}K^{ij}-K^2\right\rangle.
\end{equation}
In order to make $\langle R\rangle\approx 0$, we must have
\begin{equation}
\Lambda\approx -\frac{1}{2}\left\langle K_{ij}K^{ij}-K^2\right\rangle.
\end{equation}
Expanding the terms $K_{ij}K^{ij}-K^2$ we obtain
\begin{eqnarray}\label{kab-ksquare}
&&K_{ij}K^{ij}-K^2 \\
=&&\sum_{i\neq j\neq k}M_kK_{ij}^2+\sum_{\{i, j\}\neq\{k, l\}}\left(g^{ik}h^{jl}-g^{ij}g^{kl}\right)K_{ij}K_{kl}, \nonumber
\end{eqnarray}
where
\begin{equation}
M_k=g^{ii}g^{jj}-\left(g^{ij}\right)^2, \quad k\neq i\neq j,
\end{equation}
is the $k$th principal minor of $g^{ij}$. Since by definition the metric matrix $g^{ij}$ is positive definite, we have $M_k>0$.

According to \cite{PhysRevLett.123.131302}, $K_{ij}$ and $-K_{ij}$ are equally likely, thus, for $\{i, j\}\neq\{k,l\}$, the following four pairs of components
\begin{equation}
(K_{ij}, K_{kl}),\,(K_{ij}, -K_{kl}),\,(-K_{ij}, K_{kl}),\,(-K_{ij}, -K_{kl}) \nonumber
\end{equation}
are also equally likely. Then because in general, there is no particular relationship between the components of the extrinsic curvature, we have, for the second term in \eqref{kab-ksquare}, the above four cases would statistically cancel each other that the macroscopic spatial average
\begin{equation}\label{zero macroscpic average}
\left\langle\left(g^{ik}g^{jl}-g^{ij}g^{kl}\right)K_{ij}K_{kl}\right\rangle=0, \quad \{i, j\}\neq\{k, l\}.
\end{equation}
So only the first term in \eqref{kab-ksquare} survives after the spatial averaging that we have
\begin{equation}\label{negative lambda}
\Lambda\approx -\sum_{\substack{1\leq i<j\leq 3\\i\neq j\neq k}}\langle M_kK_{ij}^2\rangle<0.
\end{equation}

The cosmological constant $\Lambda$ in \cite{PhysRevLett.123.131302} is generated by quantum fields vacuum fluctuations. Its sign and magnitude depend on the exact particle content of the Universe, while the right hand side of \eqref{negative lambda} depends on the Planck scale gravity fluctuations. There is no known physical mechanism to make them equal. So it is necessary to introduce the bare cosmological constant $\Lambda_B$ so that we can adjust $\Lambda_B$ to guarantee a negative cosmological constant and the following equation is satisfied to make $\langle R\rangle$ small:
\begin{equation}
\Lambda_{\mathrm{eff}}=\Lambda_B+\Lambda\approx -\sum_{\substack{1\leq i<j\leq 3\\i\neq j\neq k}}\langle M_kK_{ij}^2\rangle<0.
\end{equation}
This approach to tackle the cosmological constant problem has been presented in \cite{Qingdi:2019, Wang:2019mee}.

More interestingly, the cosmological constant $\Lambda$ generated by the quantum fields vacuum fluctuations is actually not a constant. $\Lambda$ itself also fluctuates since the vacuum state is not an eigenstate of the stress energy tensor operator and the magnitude of this fluctuation is as large as $\Lambda$ itself. This would lead to a parametric resonance effect on the oscillation of the local scale factor $a$ and the oscillation amplitude of $a$ would grow exponentially. When $\Lambda_B$ is dominant over $\Lambda$, the fluctuation in $\Lambda$ serves as a perturbation on the Planck scale spacetime fluctuations and the parametric resonance effect is weak. This leads to the slowly accelerating expansion of the Universe which may explain the mysterious ``dark energy". In this scenario, no fine-tuning of the bare cosmological constant $\Lambda_B$ is needed. See more details about how this effect of fluctuations in $\Lambda$ gives a small effective cosmological constant in references \cite{Qingdi:2019, Wang:2019mee}.

On the other hand, the phase transitions in the early Universe may effectively shift $\Lambda_B$ to values comparable to the fluctuations of $\Lambda$. In this case the parametric resonance becomes strong and thus be able to produce the inflation. This might explain the ``inflaton field" which deserves further studies.

\section{Conclusion}
We have shown that the idea of hiding the cosmological constant in Planck scale curvature fluctuations does not work in the way proposed in \cite{PhysRevLett.123.131302}. The hiding is just an illusion maintained by the external forces acting on the Eulerian observers. In particular, it does not work for a positive cosmological constant due to the universal divergences of the geodesics. The small scale spacetime fluctuations do not help in this situation. Fortunately, this idea works for a negative cosmological constant in a different way. Interestingly, following the argument that the initial data $K_{ij}$ and $-K_{ij}$ are equally likely, the sign of the cosmological constant just needs to be negative to make the average spatial curvature $\langle R\rangle$ small.

The essential idea of \cite{PhysRevLett.123.131302} is that we should consider the inhomogeneity of spacetime at Planck scale and the expansion and contraction of space at that scale are equally likely so that they can cancel each other to hide the cosmological constant. This idea was originally presented in \cite{PhysRevD.95.103504} and further developed in \cite{Qingdi:2019, Wang:2019mee}, although there are some crucial differences.

\bibliographystyle{unsrt}
\bibliography{how_vacuum_gravitates}

\end{document}